\documentclass{esannV2}
\usepackage[dvips]{graphicx}
\usepackage[utf8]{inputenc}
\usepackage{amssymb,amsmath,array}
\usepackage[a4paper, top=5.2cm, bottom=5.2cm, left=4.4cm, right=4.4cm]{geometry}
\usepackage{booktabs}
\usepackage[table]{xcolor}
\usepackage{anyfontsize}
\usepackage{subcaption}

\begin{document}
\title{Lost in Modality: Evaluating the Effectiveness of Text-Based Membership Inference Attacks on Large Multimodal Models}

\author{Ziyi Tong and Feifei Sun and Le Minh Nguyen
%
%
\vspace{.3cm}\\
%
Japan Advanced Institute of Science and Technology - Information Science \\
1-1 Asahidai, Nomi, Ishikawa 923-1292 - Japan
%
}

\maketitle

\begin{abstract}
Large Multimodal Language Models (MLLMs) are emerging as one of the foundational tools in an expanding range of applications. Consequently, understanding training-data leakage in these systems is increasingly critical. Log-probability-based membership inference attacks (MIAs) have become a widely adopted approach for assessing data exposure in large language models (LLMs), yet their effect in MLLMs remains unclear. We present the first comprehensive evaluation of extending these text-based MIA methods to multimodal settings. Our experiments under vision-and-text (V+T) and text-only (T-only) conditions across the DeepSeek-VL and InternVL model families show that in in-distribution settings, logit-based MIAs perform comparably across configurations, with a slight V+T advantage. Conversely, in out-of-distribution settings, visual inputs act as regularizers, effectively masking membership signals.
\end{abstract}

\section{Introduction}

\begin{figure}[h!]
\centering
\includegraphics[scale=0.085]{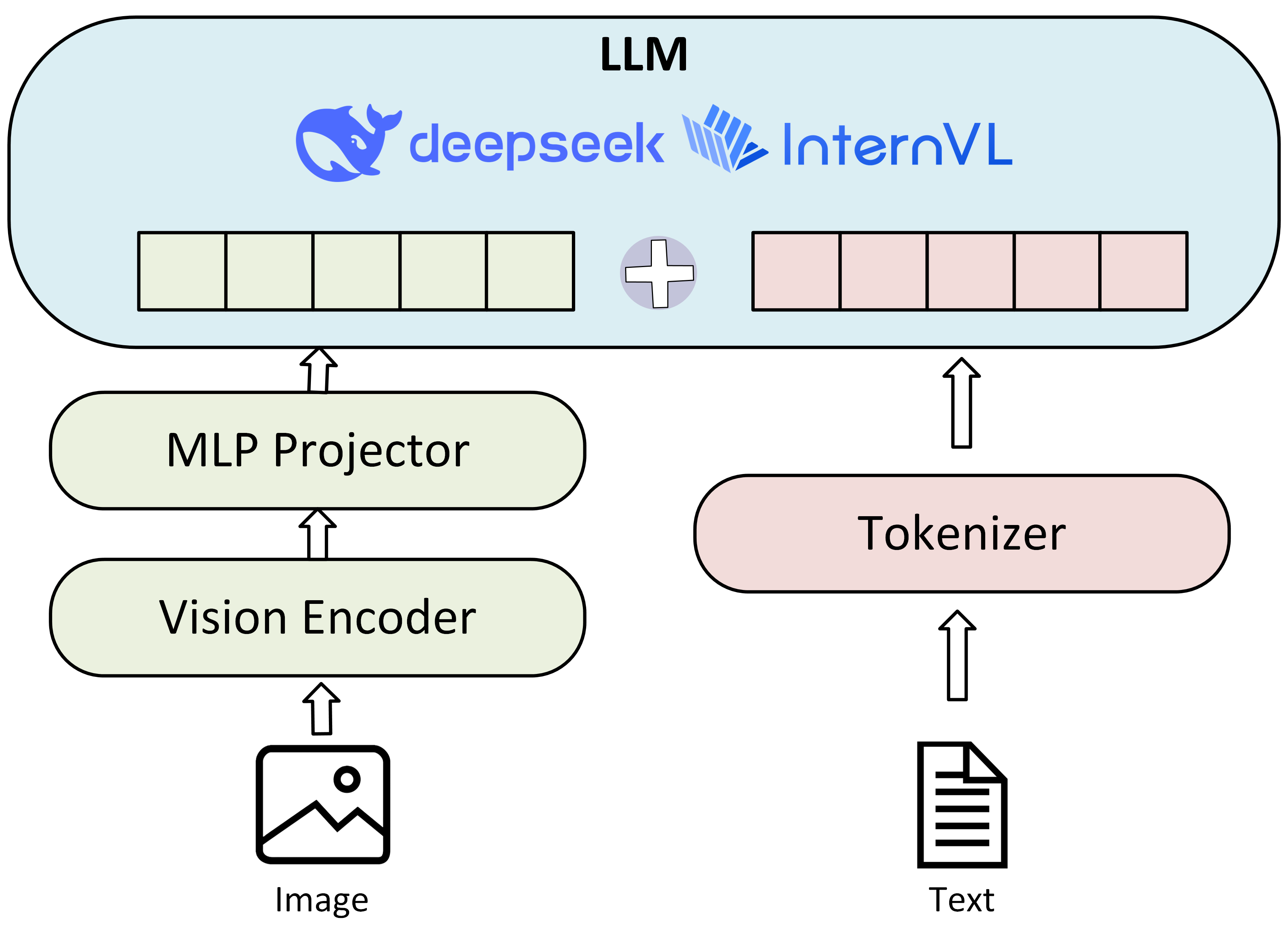}
\caption{Architecture schematic of the vision–language fusion pipeline in modern MLLMs.}\label{Fig:vllm_schematic}
\end{figure}

LLMs and MLLMs have progressed rapidly \cite{lu2024deepseek,wu2024deepseek,chen2024far,chen2024expanding}, heightening concerns about training-data exposure and motivating research on membership inference attacks\cite{tong2025pretraining}. For text input, logit-based MIA on LLM has advanced substantially, with recent methods achieving AUROC values near 90\%\cite{xie2024recall}. MLLMs integrate a vision encoder with a language decoder (Figure~\ref{Fig:vllm_schematic}), where visual embeddings are projected into the language model’s representation space and fused with text representations to condition next-token prediction\cite{li2024membership}. Although ground-truth image tokens are not available, the text logits remain conditioned on visual embeddings, which preserves the applicability of logit-based MIA.

However, it remains unclear whether these state-of-the-art, text-targeted MIA methods can be reliably applied to multimodal architectures, or how vision-conditioned text representations respond to logit-based membership inference. Addressing these questions is essential for assessing the privacy risks of modern MLLMs. 

To address these questions, we conduct experiments on four MLLMs under both in-distribution(ID) and out-of-distribution(OOD) conditions, using text-only and multimodal inputs to isolate the influence of visual features on membership leakage. Experimental results show that Visual inputs suppress MIA signals in OOD settings and that the impact of MIA is highly model-dependent.

Our contributions are as follows:
First, we offer a comprehensive evaluation of state-of-the-art text-based MIA methods on multimodal inputs, revealing how membership signals behave when visual and textual streams interact.
Second, we conduct a cross-model comparison showing that visual features modulate MIA effectiveness in model-dependent ways, driven by differing vision-text interactions.
Third, we analyze in-distribution and out-of-distribution settings, demonstrating that natural distribution shifts have a substantial impact on membership inference, often overshadowing the effect of multimodal fusion.

\section{Related Work}

\textbf{Membership Inference Attack.}  Membership inference is an attack method in which an adversary attempts to determine whether a specific data sample was included in a model’s training set\cite{tong2025pretraining}. Early text-targeted approaches use input loss directly \cite{yeom2018privacy} or calibrate it with a reference model\cite{carlini2022membership}. Compression-based variants such as \cite{carlini2021extracting} compress token-loss sequences to smooth high-variance noise and expose a lower-entropy signal useful for membership inference. More recent techniques leverage minimum token probabilities, including \cite{shi2023detecting} and its normalized extension \cite{zhang2024min}. \cite{xie2024recall} measures membership by comparing conditional and unconditional likelihoods under non-member prefixes. Image-targeted MIAs have also emerged. \cite{li2024membership} utilizes the Rényi entropy of next-token distributions as a robust confidence signal.

\textbf{Multimodal Large Language Models.} Multimodal large language models (MLLMs) extend language models by enabling joint reasoning over images and text. A widely adopted fusion paradigm maps the vision encoder’s outputs into the language model’s embedding space using a lightweight MLP projection adapter, allowing the LLM to process visual features as part of its token sequence. Recent architectures follow this design pattern, including DeepSeek-VL\cite{lu2024deepseek} and its MoE-based successor DeepSeek-VL2\cite{wu2024deepseek}, which focus on general-purpose reasoning, as well as InternVL-1.5\cite{chen2024far} and InternVL-2.5\cite{chen2024expanding}, which incorporate high-resolution vision encoders to enhance perceptual grounding. Despite their rapid progress, the privacy characteristics of these MLLMs remain largely unexplored.

\section{Experiments}
\subsection{Problem Setting}
Let $M$ be an autoregressive multimodal language model that produces a next-token distribution conditioned on an input sequence. Each VQA sample is represented as 
\[
x = (\mathbf{v}, \mathbf{q}, \mathbf{c}),
\]
where $\mathbf{v}$ is the image, $\mathbf{q}$ the question, and $\mathbf{c}$ the candidate answers. Let $D$ denote the dataset used to train $M$. The goal of membership inference is to determine, for a target sample $x$, whether $x \in D$ or $x \notin D$. 

Under a gray-box setting, the adversary has access to the model's token-level logits and computes a membership score $S(x; M)$ from these logits. The score is then thresholded to predict membership:
\[
m(x) = 
\begin{cases}
1 & \text{if } S(x; M) \ge \tau,\\[0.3em]
0 & \text{otherwise},
\end{cases}
\]
where $m(x)=1$ indicates that $x$ is a member. Our experiments evaluate the effectiveness of various logit-based MIA methods in distinguishing members from non-members in multimodal VQA settings.

\subsection{Models}

MLLMs typically pair a vision encoder with an LLM and fuse modalities by projecting visual features into the LLM embedding space through an MLP adapter. To compare architectures that emphasize different components, we evaluate four representative models: DeepSeek-VL(\textit{deepseek-vl-7b-base})\cite{lu2024deepseek} and DeepSeek-VL2(\textit{deepseek-vl2-small})\cite{wu2024deepseek}, which prioritize semantic reasoning with an MoE-enhanced decoder, and InternVL-1.5(\textit{Mini-InternVL-Chat-4B-V1-5})\cite{chen2024far} and InternVL-2.5(\textit{InternVL2\_5-4B}) \cite{chen2024expanding}, which employ high-resolution vision encoders to strengthen perceptual grounding.

\subsection{Datasets}

A major challenge in MIA research is obtaining member and non-member samples, since large-scale pretraining often obscures provenance\cite{tong2025pretraining}. ScienceQA\cite{lu2022learn} provides a rare case with explicit documentation: both DeepSeek-VL\cite{lu2024deepseek} and InternVL1.5/2.5\cite{chen2024far,chen2024expanding} report using ScienceQA for training. DeepSeek-VL also report using ScienceQA for evaluation, and DeepSeek-VL2 reports inheriting the general VQA training data of DeepSeek-VL. We therefore treat ScienceQA-train as member data and ScienceQA-test as in-distribution non-members. To further validate robustness, we also include AI2D\cite{kembhavi2016diagram}, a VQA dataset from the same scientific domain but not reported as training data for any of the four models, forming an OOD non-member split. Together, ScienceQA-train vs. ScienceQA-test and ScienceQA-train vs. AI2D instantiate our ID and OOD settings.

\subsection{Methods}

We compare six state-of-the-art logit-based MIA methods: Loss\cite{yeom2018privacy}, Reference\cite{carlini2022membership}, Min-K\%\cite{shi2023detecting}, Min-K\%++\cite{zhang2024min}, ReCALL\cite{xie2024recall}, and Zlib\cite{carlini2021extracting}, covering loss-based, calibrated, entropy-based, and compression-based paradigms. To disentangle the contributions of text and vision, we evaluate two inference configurations:
T-only (question + choices, image masked) , and V+T (image + question + choices). We report the area under the ROC curve (AUC) as the primary metric, following standard MIA evaluation practice\cite{shi2023detecting,xie2024recall}.

\section{Results and Discussion}
As show in Table~\ref{tab:main_results}, we evaluate six MIA methods across four multimodal models in both in-distribution and out-of-distribution settings.

\newcommand{\up}[1]{\textcolor{green!60!black}{#1\,\(\blacktriangle\)}}
\newcommand{\down}[1]{\textcolor{red!70!black}{#1\,\(\blacktriangledown\)}}
\newcommand{\neutral}[1]{#1}

\begin{table}[htbp]
  \centering
  \fontsize{6}{7.2}\selectfont
  \caption{Comparison of MIA performance across models, input modalities, and datasets.}
  \begin{tabular}{lcccccc}
  \toprule
  & \multicolumn{3}{c}{\textbf{In-Distribution (ScienceQA-Test)}} 
  & \multicolumn{3}{c}{\textbf{OOD (AI2D)}} \\
  \cmidrule(lr){2-4} \cmidrule(lr){5-7}
  & \textbf{Text-only} & \textbf{V+T} & \textbf{$\Delta$} 
  & \textbf{Text-only} & \textbf{V+T} & \textbf{$\Delta$} \\
  \midrule
  
  \multicolumn{7}{c}{\textbf{DeepSeek-VL}} \\
  \midrule
  Loss          & 0.475 & 0.487 & \up{0.012}
                & 0.666 & 0.655 & \down{-0.010} \\
  Reference     & 0.525 & 0.516 & \down{-0.009}
                & 0.567 & 0.657 & \up{0.090} \\
  Zlib          & 0.487 & 0.491 & \up{0.004}
                & 0.604 & 0.603 & \down{-0.001} \\
  Min-K\%       & 0.482 & 0.502 & \up{0.020}
                & 0.715 & 0.409 & \down{-0.306} \\
  Min-K\%++     & 0.475 & 0.504 & \up{0.029}
                & 0.540 & 0.330 & \down{-0.210} \\
  ReCall        & 0.513 & 0.512 & \down{-0.002}
                & 0.442 & 0.295 & \down{-0.147} \\
  \midrule

  \multicolumn{7}{c}{\textbf{DeepSeek-VL2}} \\
  \midrule
  Loss          & 0.475 & 0.483 & \up{0.008}
                & 0.563 & 0.538 & \down{-0.025} \\
  Reference     & 0.502 & 0.493 & \down{-0.009}
                & 0.322 & 0.559 & \up{0.236} \\
  Zlib          & 0.484 & 0.490 & \up{0.005}
                & 0.563 & 0.550 & \down{-0.013} \\
  Min-K\%       & 0.479 & 0.507 & \up{0.028}
                & 0.475 & 0.313 & \down{-0.163} \\
  Min-K\%++     & 0.487 & 0.479 & \down{-0.009}
                & 0.436 & 0.305 & \down{-0.131} \\
  ReCall        & 0.500 & 0.530 & \up{0.030}
                & 0.635 & 0.591 & \down{-0.044} \\
  \midrule

  \multicolumn{7}{c}{\textbf{InternVL1.5}} \\
  \midrule
  Loss          & 0.515 & 0.503 & \down{-0.012}
                & 0.680 & 0.618 & \down{-0.062} \\
  Reference     & 0.485 & 0.475 & \down{-0.010}
                & 0.353 & 0.311 & \down{-0.042} \\
  Zlib          & 0.505 & 0.500 & \down{-0.005}
                & 0.639 & 0.605 & \down{-0.034} \\
  Min-K\%       & 0.489 & 0.500 & \up{0.011}
                & 0.501 & 0.286 & \down{-0.215} \\
  Min-K\%++     & 0.473 & 0.482 & \up{0.009}
                & 0.311 & 0.317 & \up{0.006} \\
  ReCall        & 0.516 & 0.512 & \down{-0.005}
                & 0.546 & 0.473 & \down{-0.073} \\
  \midrule

  \multicolumn{7}{c}{\textbf{InternVL2.5}} \\
  \midrule
  Loss          & 0.516 & 0.507 & \down{-0.008}
                & 0.674 & 0.593 & \down{-0.081} \\
  Reference     & 0.503 & 0.505 & \up{0.003}
                & 0.594 & 0.423 & \down{-0.171} \\
  Zlib          & 0.513 & 0.506 & \down{-0.008}
                & 0.672 & 0.594 & \down{-0.078} \\
  Min-K\%       & 0.494 & 0.484 & \down{-0.010}
                & 0.555 & 0.383 & \down{-0.172} \\
  Min-K\%++     & 0.496 & 0.493 & \down{-0.003}
                & 0.515 & 0.383 & \down{-0.133} \\
  ReCall        & 0.516 & 0.504 & \down{-0.012}
                & 0.605 & 0.561 & \down{-0.044} \\
  
  \bottomrule
  \end{tabular}
  \label{tab:main_results}
  \vspace{2pt}
{\footnotesize $\Delta = \text{V+T} - \text{Text-only}$; 
\textcolor{green!60!black}{$\blacktriangle$} indicates improvement, 
\textcolor{red!70!black}{$\blacktriangledown$} indicates degradation.}
\end{table}
In the text-only (T-only) setting, all models exhibit low susceptibility to membership inference, with AUROC scores consistently hovering near the random baseline(0.5). This aligns with prior research \cite{xie2024recall} indicating that in-distribution membership inference is still very difficult for well-generalized large language models, as the loss gap between members and non-members is minimal.

When introducing visual inputs (V+T) in the in-distribution setting, we observe minor changes in MIA performance. DeepSeek-VL and DeepSeek-VL2 show deltas mostly within ±0.03, indicating negligible impact from visual conditioning. InternVL models yield slightly more consistent gains, particularly for Min-K\% variants, but the improvements remain small. Overall, visual features do not meaningfully increase membership leakage in the in-distribution setting, as the T-only baseline is already near 0.5. 

\begin{figure}[h!]
\centering

\begin{subfigure}[b]{0.49\textwidth}
    \centering
    \includegraphics[width=\textwidth]{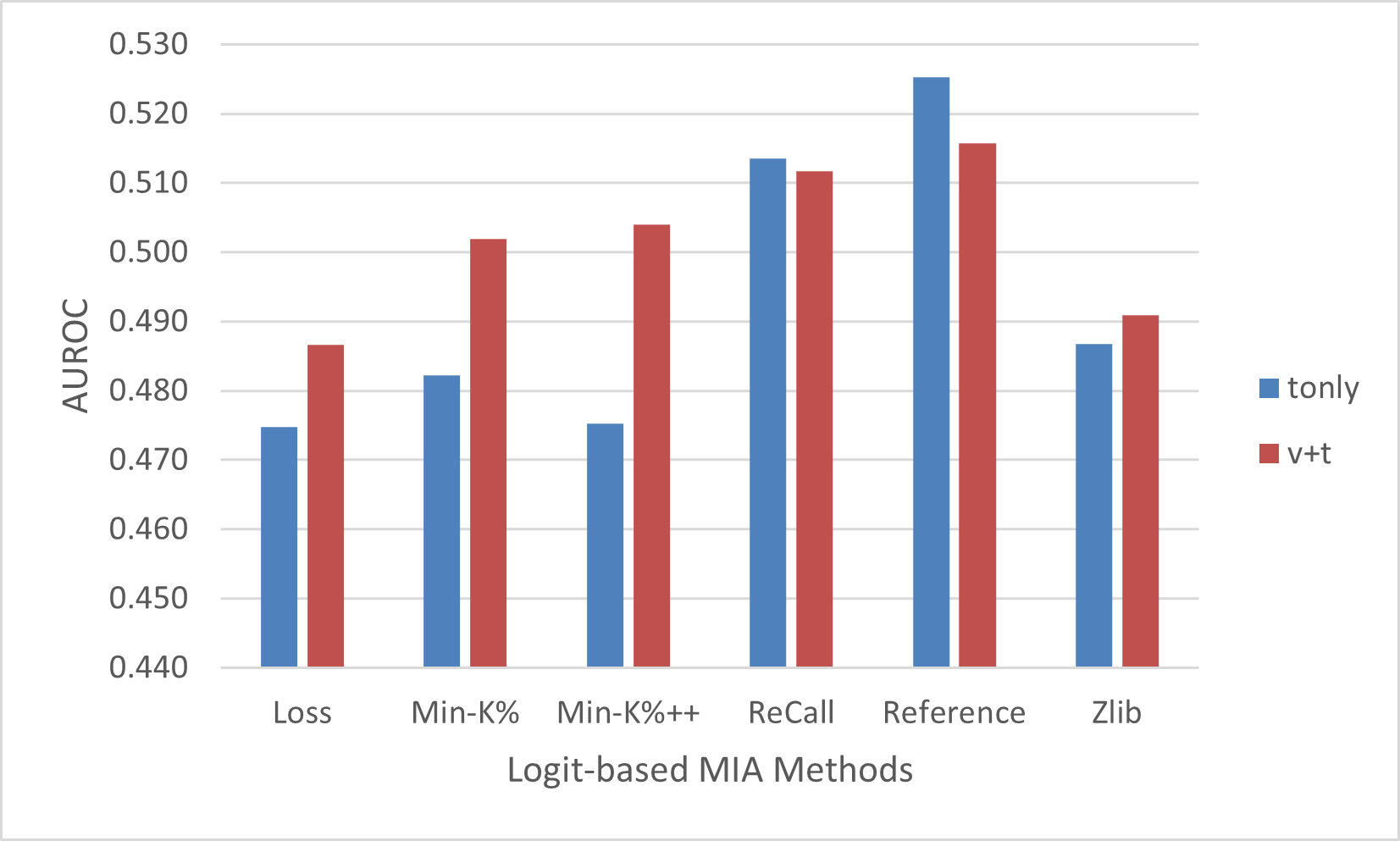}
    \caption{In distribution Setting}
    \label{fig:deepseekvl-in}
\end{subfigure}
\hfill
\begin{subfigure}[b]{0.49\textwidth}
    \centering
    \includegraphics[width=\textwidth]{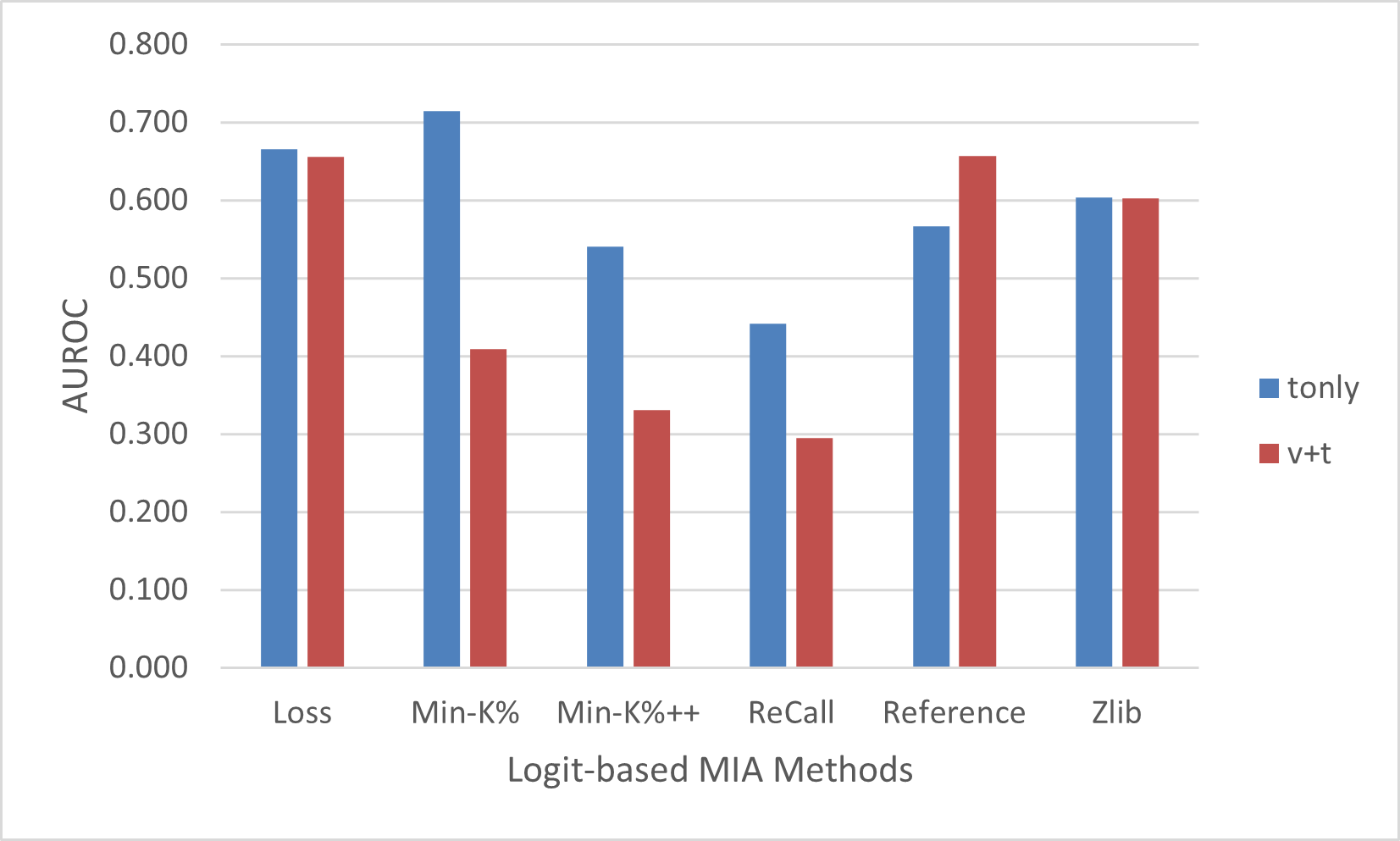}
    \caption{Out of Distribution Setting}
    \label{fig:deepseekvl-ood}
\end{subfigure}

\caption{DeepSeek-VL MIA performance in OOD (left) and In-Distribution (right) settings under text only(T-only) and vision plus text modes(V+T).}
\label{fig:deepseekvl}
\end{figure}

Conversely, in the OOD setting (AI2D), the introduction of visual inputs (V+T) causes a catastrophic drop in MIA performance. As show in Figure~\ref{fig:deepseekvl}, DeepSeek-VL sees a Min-K\%++ drop of over 0.17, and ReCall drops by 0.21. This indicates that the visual modality introduces a domain shift that masks membership signals, effectively acting as a regularizer against standard loss-based attacks.


We attribute the observed ID–OOD discrepancy to two potential factors. First, the OOD dataset (AI2D) consists of clean, schematic diagrams with lower visual complexity than the mixed-content images in ScienceQA, which may act as a confounder by masking membership signals. Second, data contamination during pre-training is a plausible alternative explanation: given the scale of web-crawled corpora, the ScienceQA test split may have been partially exposed to the models, compressing the loss gap between members and non-members. Together, these factors may explain the near-random MIA performance in the in-distribution setting and the contrast with OOD results.

\section{Conclusion}
We present the first systematic evaluation of state-of-the-art text-based MIA methods on multimodal inputs. Our results show that visual inputs can invert membership signals, that ScienceQA may suffer from data contamination, and that different attack methods exhibit varying robustness. These findings highlight important limitations of applying text-only MIA techniques to multimodal models and point to the need for multimodal-aware approaches to evaluating training-data exposure.
%





\begin{footnotesize}

\bibliographystyle{unsrt}
\bibliography{reference}

\end{footnotesize}


\end{document}